\documentclass[12pt]{article}
\usepackage{epsfig}
\def\lsim{\mathrel{\rlap {\raise.5ex\hbox{$ < $}}
{\lower.5ex\hbox{$\sim$}}}}
\def\gsim{\mathrel{\rlap {\raise.5ex\hbox{$ > $}}
{\lower.5ex\hbox{$\sim$}}}}
\newcommand{\ba}{\begin{eqnarray}}
\newcommand{\ea}{\end{eqnarray}}
\newcommand{\be}{\begin{equation}}
\newcommand{\ee}{\end{equation}}
\newcommand{\nn}{\nonumber}

\def\ra{\rightarrow}

\newcommand{\al}{\alpha_L}
\newcommand{\ar}{\alpha_R}

\renewcommand{\b}[1]{\textbf{#1}}
\renewcommand{\ell}[0]{{L}}
\renewcommand{\b}[1]{\textbf{#1}}

\begin{document}
\begin{titlepage}
\begin{flushright}
CPHT-S-093-12-00\\
hep-ph/0012255
 {\hskip.5cm}\\
\end{flushright}
\begin{centering}
\vspace{.3in}
{\bf A Pati--Salam model from branes  }   \\
   \vspace{2 cm}
{G.K. Leontaris$^1$ and J. Rizos$^{2\,,*}$} \\ \vskip 1cm
$^{1}${\it {Physics Department, University of Ioannina\\
Ioannina, GR45110, Greece}}\\
\vspace*{.5cm}
$^{2}${\it
Centre de Physique Th{\'e}orique, Ecole Polytechnique},\\
{\it F-91128 Palaiseau, France} \\
\vspace{1.5cm}
{\bf Abstract}\\
\end{centering}
\vspace{.1in}
We explore  the possibility of  embedding the Pati--Salam
 model in the context of Type I brane models.
We study a generic model with ${U(4)}_C\times{U(2)}_L\times{U(2)}_R$
gauge symmetry and matter fields compatible with a Type I brane configuration.
Examining the anomaly cancellation conditions of the surplus abelian
symmetries  we find an alternative hypercharge embedding
that is compatible with  a low string/brane scale of the order of
$5-7\,$ TeV, when the ${U(4)}_C$ and ${U(2)}_R$ brane stack
couplings are equal.
Proton stability is assured as baryon number is associated
to a global symmetry remnant of the broken abelian factors.
It is also shown that this scenario can accommodate an extra
low energy abelian symmetry that can be associated to lepton
number. The issue of fermion and especially neutrino masses is also discussed.
\\
 \vspace{0.2cm}
\begin{flushleft}
\end{flushleft}
\vfil
\hrule width 8.9cm
\vspace*{0.1cm}
$^*$ {\it On leave from Physics Department, University of Ioannina, GR45110 Ioannina, Greece}
\end{titlepage}
\section{Introduction}
It has been recently realized that in Type I  string theories
the string scale is not necessarily of the order of the  Planck mass, as it
happens in the case of heterotic models, but it can be much lower
depending on the compactification volume \cite{witten}.
Furthermore, the discovery of D-branes \cite{DBRANE},
 solitonic objects of Type I string theory,
has revolutionized the string-theory viewpoint of our world.
This includes the possibility that we are living on a $p-$dimensional
 hyper-surface,  a $D{p-1}$ brane embedded in the 10-dimensional string theory.
 The rest, $10-p$ transverse dimensions constitute the so called
 bulk space. The gauge interactions, mediated  by open strings,
 restrict their action in the brane, while gravitational interactions,
 mediated by closed strings, can propagate in the full 10-dimensional theory.
These developments  have reinforced   expectations  that
some string radii can be brought down to the TeV range \cite{LUS}, energy
accessible to the future accelerators, and that
string theory could account for the  stabilization of   hierarchy
without invoking  supersymmetry \cite{LUS2}.

Furthermore, new techniques have been  developed for the construction of
Type I models \cite{TIS}, including the D-brane configurations, based
on Type IIB  orientifolds \cite{TIC}.
Various models,  basically variations of the Standard Model or it's
left-right symmetric extensions, have been constructed \cite{BMS1,BMS2},
using these methods.
Although some of these models are
characterised as semi-realistic, from the phenomenological point of view,
the structure of Type I string vacua  is
very rich to permit a complete classification. Hence,
 model building endeavour  needs to be carried on  until we reach
a  phenomenologically  satisfactory vacuum.

One is  tempted to  adopt a bottom-up approach \cite{BU1,BU2},
that is, to search for effective low
energy models compatible with low unification and check  their
generic phenomenological properties
\cite{BMP,bulknu,bulknulr,bulknur} and the minimal conditions
for phenomenological viability,
before proceeding to
explicit realizations in the context of string theory.
Low scale unified  models
based on gauge symmetries beyond that of the Standard Model (SM)
face several problems. Proton decay is
usually the most serious obstruction when lowering the unification mass
below the traditional grand unified scale $M_{GUT}\sim 10^{16}$
GeV, due to the existence of gauge-mediated baryon number violating
dimension-six operators.
 In addition, one needs to  understand how a rapid convergence
of the gauge couplings can occur in an energy region much shorter
than the traditional  $M_Z-M_{GUT}$ of old unified
models~\cite{Lan}.

Rapid gauge-boson mediated proton decay, excludes a wide range of gauge
groups beyond the SM, however examples of models which can in
principle avoid  this problem do exist.
A natural candidate is the Pati--Salam (PS) model~\cite{PS}, originally
proposed as a model of low unification scale.
This model has been successfully reproduced and studied  in the
context of heterotic string theory \cite{ALR}.

With regard to the  problem of coupling unification,
 there are various proposals in the literature ~\cite{LUS2,DG,BU1}.
One possibility is to assume   power law
running of the gauge couplings~\cite{DG} and obtain full coupling
unification at a low scale.
An alternative scenario is based on the observation that the
different collections of
D-branes (associated with the extended  gauge group factors)
have not necessarily equal gauge couplings. The low energy
electroweak data could then be reproduced by considering the usual
logarithmic coupling evolution while assuming equality of two (instead of
three) gauge couplings at the string scale \cite{BU1}.

In this work, we search for   D-brane
configurations  where the left-right PS  gauge
symmetry is embedded. Since supersymmetry can be broken at the
string/brane level \cite{SUSYB} we are going
 to explore  non-supersymmetric  versions of the Pati--Salam  model.
We derive a generic D-brane configuration fermion and higgs spectrum
and show that  all
the SM particles and the necessary Yukawa couplings for fermion
masses are present. We address the problems of anomaly cancellation,
hypercharge embedding,  proton decay and
gauge coupling unification. Our analysis shows that
all these problems  find  natural solutions and that
the non-supersymmetric Pati--Salam model
is compatible with intermediate/low scale D-brane scenarios.

\section{Particle assignment}

A single D-brane  carries a $U(1)$ gauge symmetry which is the
result of the reduction of the ten-dimensional Yang-Mills theory.
Therefore, a stack of $n$ parallel, almost coincident D-branes gives
rise to a $U(n)$ gauge theory where the gauge bosons correspond to
open strings having both their ends attached to some  of the
branes of the stack.
 For the embedding of the PS model we consider brane configurations
of three different stacks containing 4-2-2 branes
respectively, which give rise to a ${U(4)}_C\times{U(2)}_L\times{U(2)}_R$
 or equivalently ${SU(4)}_C\times{SU(2)}_L\times{SU(2)}_R
\times{U(1)}_C\times{U(1)}_L\times{U(1)}_R$ gauge symmetry.
\begin{figure}[!ht]
\center
\epsfig{file=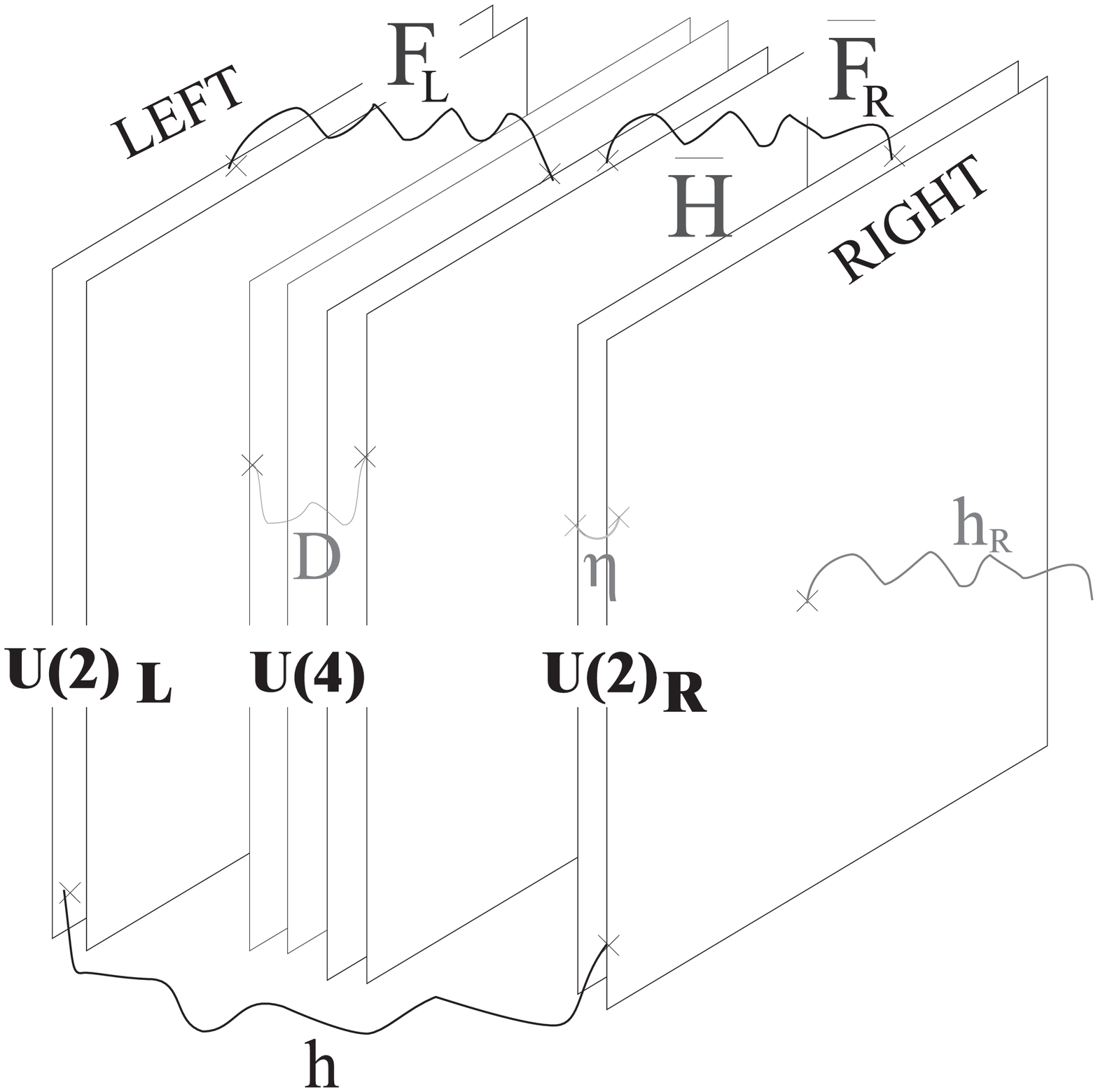,width=7cm}
\caption{\label{bfig}\it
Assignment of the Standard model particles in a
D-brane scenario with gauge group $U(4)_C\times {U(2)}_L\times
{U(2)}_R$. The standard model particles
 are assigned to  $F_L=Q+L$, $\bar{F}_R=u^c+d^c+e^c+\nu^c$ and
the electroweak Higgs to
$h=H_u+H_d$. They are all represented by strings having both their
ends attached to two different branes.
The PS breaking Higgs scalars ($\bar{H}$) are similar to $\bar{F}_R$.
In gray we represent  particles whose presence is not required in all
versions of the model. These are the extra scalar triplets
$D=\tilde{d}+\tilde{d}^c$, the right-handed doublets $h_R$ and the
singlet $\eta$.}
\end{figure}

Following the pictorial representation of Fig. \ref{bfig}
it is not difficult to see that the possible states arising from strings
with both their ends on two distinct sets
of branes can accommodate the fermions of the SM as
well as the necessary Higgs particles to break the gauge symmetry.
For example, an
open string with one end on the $U(4)$ brane and the other end on
the $U(2)_L$ brane transforms as $(\b{4},\b{2}_L)$ whilst is a singlet
under $SU(2)_R$. Thus, under the PS group  the corresponding state
is written as $(\b{4},\b{2},\b{1})$.
 Due to the decompositions under the chains $U(n)\ra SU(n)\times U(1)$
$(n=4,2,2)$ all such states carry charges under three surplus
$U(1)$ factors. Normalizing appropriately
\footnote{We assume the $U(n)\sim SU(n)\times U(1)$  generators
$T_a, a=1,\dots,n^2$ to be normalized as
${\rm tr}\,T_a T_b=\frac{1}{2}\delta_{ab}$ and the $SU(n)$
coupling constant to be $\sqrt{2 n}$ times the $U(1)$ coupling constant.}
these charges are
$+1,-1$ for the vector/vector-bar  representation of $SU(n)$,
 and thus, the standard model particle  assignments are
\ba
F_L&=&(\b{4},\b{2},\b{1},+1,\al,0)
=Q({\bf3},{\bf2},\frac 16) + \ell{({\bf1},{\bf2},-\frac 12)}\nonumber\\
\bar{F}_R&=&(\bar{\b{4}},\b{1},\b{2},-1,0,\ar)
=u^c({\bf\bar 3},
{\bf1},-\frac 23)+d^c({\bf\bar3},{\bf1},\frac 13)+
e^c({\bf1}, {\bf1},1)+ \nu^c({\bf1},{\bf1},0)
\label{fermions}
\ea
where  $\al=\pm1,\ar=\pm1$ depending on the ${U(1)}_L, {U(1)}_R$
charges  of $\b{2}_L, \b{2}_R$.
The electroweak breaking scalar doublets can arise from the bi-doublet
\ba
h&=&(\b{1},\b{2},\b{2},0,-\al,-\ar) =H_u(\b1,\b2,+\frac{1}{2})+
H_d(\b1,\b2,-\frac{1}{2})\label{ehiggs}
\ea
where we have chosen the  $U(1)_{L,R}$  charges so that the Yukawa term
$F_L\bar{F}_R h$ which provides with masses all fermions, is allowed.
The PS breaking Higgs scalar particles are
\ba
\bar{H}
&=&(\bar{\b{4}},\b{1},\b{2},-1,0,\gamma)=
u^c_H({\bf\bar 3},
{\bf1},-\frac 23)+d^c_H({\bf\bar3},{\bf1},\frac 13)+
e^c_H({\bf1}, {\bf1},1)+ \nu^c_H({\bf1},{\bf1},0)\,.
\label{GUThiggs}
\ea
Without loss of generality we can choose  $\al=\ar=1$ which is equivalent to
 measuring left (right) $SU(2)_{L(R)}$
vector representation ${U(1)}_{L(R)}$-charges
in $\al (\ar)$ ``units" respectively.

Additional states can arise from strings having both their ends at
the same brane.  Among them one finds the $SU(4)$ sextet
\ba
D(\b{6},\b{1},\b{1},+2,0,0)={\tilde{d}^c}({\bf\bar3},{\bf1},\frac 13)+\tilde{d}({\bf3},{\bf1},-\frac 13)
\label{sext}\ea
(see Fig. 1),
which can be used to provide  masses  to the Higgs remnants (one
$d$-like triplet) of the PS breaking
 Higgs mechanism (see section \ref{pd}). Further, one may generate
a $U(1)_R$ charged singlet
\ba
\eta=(\b{1},\b{1},\b{1},0,0,+2)\label{exsin}
\ea
which, as will become clear later (see section \ref{sb}), can
be used for breaking an additional abelian symmetry.
Possible states include also strings having one end attached to a brane
and the other in the bulk while among them we find the ${SU(2)}_R$
doublet
\ba
h_R=
(\b{1},\b{1},\b{2};0,0,+1)
\label{exdoub}
\ea
which will be also used later  for an alternative  breaking of an
additional abelian symmetry.

\section{Anomalies\label{sanom}}

An essential difference between heterotic and Type I effective string theories
is the number of potentially anomalous abelian factors. In the heterotic case
only one such factor is allowed with rather tight restrictions on the form
of its mixed anomalies, due to their relation with the dilaton multiplet.
 Type I theory is more tolerant,  many anomalous abelian factors
can be present and their cancellation is achieved through a generalized
Green--Schwarz mechanism ~\cite{kk} which utilizes the axion fields of
the Ramond--Ramond sector~\cite{Ibanez:1999it,Lalak:1999bk},
providing masses to the corresponding anomalous gauge bosons.
However, in Type I  models,
unlike the heterotic string case, gauge boson masses are  fixed by
undetermined vacuum expectation values and
therefore the $U(1)$ gauge bosons may be light. Another important
characteristic of Type I abelian factors is that their breaking
leaves behind global symmetries, that can be useful for
phenomenology.

As can be seen from the fermion charge assignments  (\ref{fermions}),
 the abelian gauge group factors have mixed anomalies with
 $SU(4)$, ${SU(2)}_L$ and ${SU(2)}_R$.
{}We present these anomalies in matrix form
\ba
A=\left(\begin{array}{rcc}
0&3&3
\\
6&6
&
0
\\
-6&
0
&6
\end{array}\right)
\ea
where its lines correspond to the abelian factors ${U(1)}_C$,
${U(1)}_L$, ${U(1)}_R$ and its columns to the non-abelian groups
$SU(4)$, ${SU(2)}_L$, ${SU(2)}_R$.
 {}From the point of
view of the low energy theory, it is crucial to examine whether
there are any combinations of anomaly free abelian generators.
This would imply the  existence of additional unbroken
$U(1)$ factors at low energies which may result to interesting
phenomenology. For example, the existence of $U(1)$ factors offers
the possibility to define the hypercharge generator in various
ways provided that the fermion and electroweak breaking  Higgs
particles  acquire the standard hypercharge assignments.
We find that there exists only one
non-anomalous combination
\ba
{\cal H }= T_C-T_L+T_R
\label{NAU1}
\ea
which also has the advantage of being  free from gravitational
anomalies
as both, $\mbox{trace}({\cal H })=0$ and
${\rm trace}({\cal H}^3)=0$.

One may wonder about the existence of such additional anomaly-free
abelian symmetry (on top of $B-L$ and $Y$). The reason is that
none of the SM fermions is charged under this symmetry.
Actually, the only states potentially
charged under ${U(1)}_{\cal H}$ are the PS
breaking Higgs scalars $\bar{H}$ (and  the scalars $h_R$,
$D$, $\eta$).
Later on, we will associate the value of the parameter $\gamma$,
which determines the PS breaking Higgs charges with the symmetry
breaking pattern and  discuss the possibility of survival of $U(1)_{\cal H}$
at low energies.

Thus, at this stage, assuming  that all  anomalous
abelian combinations will break, we are left with an effective
  theory with gauge symmetry
 ${SU(4)}_C\times{SU(2)}_L\times{SU(2)}_R\times{U(1)}_{\cal H}$.

\section{Symmetry breaking and the Hypercharge generator\label{sb}}

We next analyse the pattern of symmetry breaking.
The Higgs scalar  $\bar H$, (provided that an appropriate potential
exists),
will acquire a non-zero vev and break the original symmetry down to
SM augmented by a $U(1)$
factor. Subsequently,  the electroweak symmetry breaking
occurs via non-vanishing vevs of the
$H_u,H_d$  Higgs particles. Since the bi-doublet $h$, and consequently
$H_u,H_d$ are neutral under ${U(1)}_{\cal H}$, there will be always
a leftover   abelian
combination whose structure is completely determined by
the $\bar{H}$ charge ($\gamma-1$)
(see  relations (\ref{GUThiggs},\ref{NAU1})).
Thus, the hypercharge generator will be, in general,
 a linear combination of the usual PS generator
and the additional abelian gauge factor $U(1)_{\cal H}$:
\ba
Y=\frac{1}{2}\,Q_{B-L} +\frac{1}{2}\, Q_{3R}+c\,Q_{\cal H},
\label{ynew}
\ea
where $c$ is to be determined by the symmetry breaking.
The  PS breaking Higgs particles, $\bar{H}$,
contain two  potential SM singlets
with ${U(1)}_{B-L}\times{U(1)}_{3R}\times{U(1)}_{\cal H}$
charges
\ba
N_+=\left(+1,+1,-1+\gamma\right)\ ,\ N_-=\left(+1,-1,-1+\gamma\right).
\ea
When the minimum of the scalar potential occurs for either
$N_+ = \langle e^c_H\rangle,$  or $N_- = \langle\nu^c_H\rangle$
 different than zero,  the gauge symmetry breaks to the SM times
an additional abelian factor.
Extra abelian factors, although in principle consistent with
low energy data \cite{EL},
necessitate a breaking mechanism. An interesting  property of
the model presented here is
that the appropriate scalar fields, which can break these extra abelian factors,
are naturally generated in the D-brane scenario.
These are the singlet field  $\eta$ (\ref{exsin})
and the
the right-handed doublet  (\ref{exdoub})
\ba
h_R(\b1,\b1,\b2,0,0,+1)&=&h_R^+(\b{1},\b{1},0,+1,+1)+h_R^-(\b{1},\b{1},0,-1,+1)\,.\nonumber
\ea
Depending on the value of $\gamma$ there are two possible breaking
patterns, which are presented in Table \ref{tab}.
\renewcommand{\arraystretch}{1.5}
\begin{table}
\center
\begin{tabular}{|l|c|c|c|l|c|c|}
\hline
&c&vev&$\gamma$&\multicolumn{1}{|c|}{$Y$}&additional $U(1)$&
\parbox{3cm}
{\mbox{vevs that  break}\\ {\mbox{additional $U(1)$}}}\\
\hline
1&0&${N}_-$&+1&$\frac{1}{2}\,Q_{B-L}+\frac{1}{2}\,Q_{3R}$
&$Q_{\cal H}$
&$\langle\eta\rangle$\\
\hline
2&$\frac{1}{2}$&$N_+$&-1&$\frac{1}{2}\,Q_{B-L}+\frac{1}{2}\, Q_{3R}+\frac{1}{2}\,Q_{\cal H}$
&$\frac{1}{2}\,Q_{B-L}-\frac{1}{2}\,Q_{3R}$
&$\langle h_R^-\rangle$\\
\hline
\end{tabular}
\caption{\label{tab}\it The two  symmetry breaking patterns
of $SU(4)\times {SU(2)}_R\times {U(1)}_{\cal H}$  and the corresponding
PS Higgs vevs, $N_+,{N}_-$, their right
chirality $(\gamma=\pm 1)$,
the resulting hypercharge generator, the leftover abelian factor
and the scalar fields that can break this extra abelian factor.}
\end{table}
The ${U(1)}_{\cal H}$-charge of the Higgs field $\bar H$
in the  two  cases
is  zero and  $-2$ respectively.
{}For  $\gamma=+1$ (case 1 in Table \ref{tab})
 assuming  a non-zero vev for $N_-$,
the surviving abelian factors are of the
form $\frac{1}{2}\,Q_{B-L}+\frac{1}{2}\,Q_{3R}+z\,Q_{\cal H}$ (where $z$ is
an arbitrary parameter). We are free to choose the hypercharge
generator as traditionally (putting $z=0$)
and leave  $\cal H$ as the surplus abelian
factor. This is also dictated by the fact that the additional
Higgs field $\eta$ has the right charges
to break  completely ${U(1)}_{\cal H}$. The last breaking can in
principle happen to a scale which can be lower than the
PS breaking scale and can lead to a model with an additional
$U(1)$  symmetry at low energies.
{}For $\gamma=-1$ (case 2 in Table \ref{tab}),
  provided  $N_+$ develops a vev,
 the surviving abelian
factors have the generic form
$\left(2\,z-\frac{1}{2}\right)\,Q_{B-L}+\frac{1}{2}\,Q_{3R}+z\,Q_{\cal
H}$.
Assuming  vevs for $h_R^-$, the only unbroken
 combination left is
$Y=\frac{1}{2}\,Q_{B-L}+\frac{1}{2}\,Q_{3R}+\frac{1}{2}\,Q_{\cal
H}$.
This is a novel hypercharge embedding which as discussed
earlier  does not affect  the fermion and the
electroweak Higgs  charges.
Again the extra $U(1)$ breaking scale is not necessarily
associated with the $M_R$ symmetry breaking scale but can be lower.
 The  breaking of the additional  $U(1)_{\cal H}$
 symmetry  implies the existence
of a new $Z'$-boson. This is a very interesting prediction since
recent analyses show compatibility of electroweak data
with the existence of an additional gauge
boson with  mass of a few hundred GeV~\cite{EL}.

\section{\label{sd}Gauge coupling running and the weak mixing angle}

 Our main aim in this section is to
ensure that the above constructions imply the correct values
for the weak mixing angle and the gauge couplings at $M_Z$.
Moreover, it would be of particular interest if the
present D-brane model is compatible with a
low energy unification scale. The one loop renormalization group
equations are of the form
$\frac{1}{\alpha(\mu_2)}=\frac{1}{\alpha(\mu_1)}-\frac{b_i}{2\,\pi}
\log\left(\frac{\mu_2}{\mu_1}\right)$ and in our analysis
we will assume two different energy regions $(\mu_1,\mu_2)=$
\{($M_Z$,$M_R$) , ($M_R,M_U$)\}, where $M_R$ is the $U(4)\times U(2)_R$
breaking scale and $M_U$ the string scale. For simplicity, we will
also assume
that the additional $U(1)$ breaks at the same scale as the PS
symmetry, that is
$M_R=M_{Z'}$. The beta functions are $b_3, b_2, b_Y$ for the first
interval and $b_4, b_L, b_R, b_{\cal H}$ for the second in a
self-explanatory notation. The matching conditions at $M_R$
assuming properly normalized generators (all
group generators $(T_a)\,$are normalized according to ${\rm tr}
(T_a\,T_b)=\frac{1}{2}\,\delta_{ab}$),  are $\frac{1}{\alpha_Y(M_R)}=
\frac{2}{3}\,\frac{1}{\alpha_4(M_R)}+\frac{1}{\alpha_R(M_R)}+c^2\,
\frac{1}{\alpha_{\cal H}(M_R)}$ and $\alpha_3(M_R)=\alpha_4(M_R)$.
Moreover, at $M_U$ we have $\frac{1}{\alpha_{\cal H}(M_U)}=
\frac{8}{\alpha_{4}(M_U)}+\frac{4}{\alpha_{R}(M_U)}+\frac{4}{\alpha_{L}(M_U)}$.
Solving the RGE system together with the matching conditions,
 we derive the formulae for the low energy quantities
as  functions of the brane couplings ($\alpha_4,\alpha_R,
\alpha_L$),
the beta function coefficients and the scales $M_U,M_R$:
\newcommand{\Mfunction}[1]{{#1}}
\ba
\sin^2\theta_W(M_Z)=\frac{3}{8(1+6\,c^2)}&\times&\left[1+
\frac{\alpha_{em}(M_Z)}{6\,\pi}
\left\{\vphantom{\frac{1}{6\,\pi }}\right.\right.\nonumber\\
&~&\left.\left.
     \left( -2\,
        {{\Mfunction{b}}_4} -
       3\,c^2\,{b_H} +
       \left( 5 + 48\,c^2
          \right) \,
        {{\Mfunction{b}}_L} -
       3\,{b_R} \right) \ln (\frac{\mu }
       {{M_R}})\,\right.\right.\nonumber\\
&~&\left.\left.\mbox{}+
     \left( \left( 5 +
          48\,c^2 \right) \,
        {{\Mfunction{b}}_2} -
       3\,{b_Y} \right)  \log (\frac{{M_R}}
       {{M_Z}})\right.\right.\nonumber\\
&~&\left.\left.\mbox{}- 6\,\pi\,\left( \frac{2\,
        \left( 1 +
         12\,c^2 \right) }
        {{{3\,\alpha }_4}} -
     \frac{5 + 36\,c^2}
      {{{3\,\alpha }_L}} +
     \frac{
        \left( 1 + 4\,c^2
          \right) }{{{\alpha
          }_R}}\right)\right\}\right]\label{be1}
\ea
\ba
\frac{1}{\alpha_3 (M_Z)}=\frac{3}{8(1+6\,c^2)}&
\times&\left[\frac{1}{\alpha_{em}(M_Z)}-
\frac{1}{2\,\pi}{\,
     \left( -2\,
        \left( 1 + 8\,c^2
          \right) \,
        {{\Mfunction{b}}_4} +
       c^2\,{b_H} + {b_L} +
       {{\Mfunction{b}}_R}
       \right) } \ln (\frac{\mu }
       {{M_R}})\right. \nonumber\\
&~&\left. - \frac{1}{{6\,\pi }}{\,
     \left( 3\,{b_2} -
       8\,\left( 1 +
         6\,c^2 \right) \,
        {b_3} + 3\,{b_Y}
       \right)\ln (\frac{{M_R}}
       {{M_Z}}) }\right.\nonumber\\
&~&       \left.+
  \left( 1 + 4\,c^2 \right) \,
   \left( \frac{2}
      {{{\alpha }_4}} -
     \frac{1}
      {{{\alpha }_L}} -
     \frac{1}{{{\alpha }_R}}
     \right)\right]\label{be2}
\ea
Assuming coincident brane stacks,
$\alpha_4=\alpha_R=\alpha_L$, as in the case of {\it grand unification},
  the last term in both equations
vanishes and we can calculate $M_R$ and $M_U$ using low energy data.
As expected  we obtain $M_R\sim 10^{12}, M_U\sim
10^{16} GeV$ for $c=0$ (assuming minimal matter content). The choice
$c=\frac{1}{2}$ is not possible in this case, since it requires
 $M_R<M_Z$.

As already noted, since the various groups live in different
brane-stacks, the initial values of the gauge couplings is not
necessarily the same. It is thus tempting  to explore this possibility in
order to obtain low energy string/brane scale.  However, in order not
to loose predictive power we shall
choose  two of the three (brane) couplings to be
equal. We call this scheme
 {\it petite unification}\footnote{For the introduction of this term
see \cite{PU}.}   as opposed
to  {\it grand unification} where all couplings are equal.
Thus, in the present  {\it petite unification} scenario
we end up with three distinct cases, namely $\al=\ar\ne \alpha_4$,
$\al=\alpha_4\ne \ar$ and $\ar=\alpha_4\ne \al$.

For $\alpha_R=\alpha_4$ the string/brane scale $M_U$ is given by
\ba
\log\frac{M_U}{M_Z}&=& \frac{B_1}{3\,B_2}\,\log\frac{M_R}{M_Z}\nn\\
&~&\mbox{}+{2\,\pi}\,\frac{3 ((1+4 c^2)\sin^2\theta_W-1)\alpha_3(M_Z)+
(5+36\,c^2)\,\alpha_{em}(M_Z)}{3\,B_2\,\alpha_{em}(M_Z)\,\alpha_3(M_Z)}
\ea
where
$B_1=-5\,{b_3} - 3\,{c^2}\,\left( 4\,{b_2} + 12\,{b_3} -
     12\,{b_4} + {b_H} - 4\,{b_L} \right)  +
  3\,\left( {b_4} - {b_R} + {b_Y} \right)$ and
   $B_2=- {b_R}+\left( 1 + 12\,{c^2} \right) \,{b_4} -
  {c^2}\,\left( {b_H} - 4\,{b_L} \right)  $.
The beta functions depend on the details of the model particle spectrum.
Following the analysis of section \ref{sd}, we have two possibilities
for the hypercharge embedding: (i) $c=0$ where we assume that the
number of extra singlets ($\eta$) is $n_1>0$  and the number of
right-handed doublets ($h_R$) is $n_2=0$ and (ii) $c=1/2$ where  $n_1=0$ and
$n_2>0$. Furthermore, motivated by the analysis of section
\ref{pd}, with regard to the Higgs remnant triplet masses,
 we are  going to consider two  subcases for each embedding :
 $n_6=0, n_1=1$ or  $n_6=1, n_1=1$ for the case (i)  where $n_6$ is
the number of  sextets ($D$), and
$n_6=0, n_2=1$ or $n_6=1, n_2=2$ for the case (ii).
For the minimal scenario where we have three generations and
only one PS breaking Higgs multiplet $n_H=1$,
substituting the beta functions we get
$B_1=2\,{n_6}-$ $32\,\left( 1 + c \right)^2\,
      \left( -1 + 2\,c \right)  - {n_2}/2  - {c^2}\,\left( {n_1} + 2\,{n_2} \right)$
          and
$B_2={{-{n_2}/6 -
     \left( 23 + 4\,{c^2}\,
         \left( 103 + 16\,c \right)  -
        2\,{n_6} + {n_h} \right)/3 }}-{c^2}\,
      \left( {n_1} + 2\,{n_2} - 4\,{n_h}
         \right)/3$
where $n_h$ is the number of bi-doublets ($h$).

For the case (i) which corresponds to the standard Hypercharge
embedding ($c=0$) and assuming {\it petite unification}
we can obtain various values for string scale $M_U$  depending on the
unification condition. These cases have been analyzed and the
basic results are presented in Table \ref{czunf}.
One easily concludes that in all cases $M_U\gsim10^{10} GeV$.
This embedding is thus compatible with branes but not with low
scale string scenarios.
\begin{table}
\center
\begin{tabular}{|c|c|c|c|c|}
\hline
case&\parbox[s]{2.9cm}{\center\small petite unification\\ condition}&$M_U$&$M_R$&
\parbox[s]{2.4cm}{\center\small remaining\\ coupling ratio}\\
\hline
$c=0$&$\ar=\al$&$>2\times10^{12}$&$<2\times 10^{12}$&$>0.8$\\
\hline
$c=0$&$\alpha_4=\al$&$>6.1\times 10^{9}$&$>10^2$&$>0.4$\\
\hline
$c=0$&$\alpha_4=\ar$&$>6.8\times 10^{13}$&$<6.8\times10^{13}$&$>0.8$\\
\hline
$c=\frac{1}{2}$&$\alpha_R=\al$&$ <11$&$<11$&$<.15$\\
\hline
$c=\frac{1}{2}$&$\alpha_4=\al$&$-$&$-$&$-$\\
\hline
$c=\frac{1}{2}$&$\alpha_4=\ar$&$5.1\times 10^{3}-6.5\times
10^{3}$&$10^{2}-6.5\times
10^{3}$&$0.4-0.5$\\
\hline
\end{tabular}
\caption{\label{czunf}\it Limits on the  brane scale $M_U$, the intermediate scale $M_R$
and the independent coupling ratio
for various petite unification conditions and the two  hypercharge
embeddings ($Y=\frac{1}{2}\,Q_{B-L}+\frac{1}{2}\,Q_{3R}+c\,Q_{\cal
H}$). In the calculations we have taken into account the combined limits
for
two cases of minimal spectrum
 (We examine the cases $n_H=1,\,n_6=0,1,\,\,  n_1=1$  for $c=0$, and $n_6=0, n_2=1$ or $n_6=1, n_2=2$ for $c=1/2$.
 In addition we took $n_h=3$ whenever that results depend on $n_h$)
 and incorporated  the strong coupling
 uncertainties.}
\end{table}

For the case (ii) that is  $c=1/2$, we remark that
 $B_1=0, B_2=-134/3$ for the first subcase (no sextets, one right-handed
 doublet)
and $B_1=-1, B_2=-45$ for the second
(one sextet- one right-handed doublet). Hence,
 the string scale $M_U$ depends either very weakly on $M_R$ or
it does not depend at all (at the one loop). In addition,
the string scale is independent of   the number of
bi-doublets (and thus electroweak doublets).
This is actually a consequence of the combination
of the PS symmetry with the new
hypercharge embedding (\ref{ynew}) considered here which
allows us to obtain generic results for the  string/brane
scale.
Substituting the electroweak data \cite{EWD} and
taking into account the strong coupling uncertainties
 we obtain the combined range (includes both subcases which differ slightly)
\ba
 M_U= (5.1-6.5)\, {\rm TeV} \label{ls}
\ea
The non-coinciding brane coupling ratio  depends  slightly on
$M_R$ and $n_h$
and lies in the range
\ba
\frac{\al}{\alpha_4}=0.4-0.5\label{ratio}
\ea
for  $M_Z<M_R<M_U$ and $n_h=1-3$. The absolute coupling values are
$\alpha_4=\alpha_R\sim 0.07$, $\alpha_L\sim 0.03$ so we are safely
in the perturbative regime.
In addition, the low string/brane scale obtained in (\ref{ls}) is compatible with
current
limits from four-fermion interactions \cite{FF}.
We also notice that for $c=1/2$, $\alpha_4=\alpha_L$ is impossible
(since at $\mu\sim 10^{10}$GeV  $\ar$ develops negative values)
 while $\alpha_R=\alpha_L$
yields a unification scale of 7 GeV, which is obviously excluded.
The above results are also summarized in Table \ref{czunf}.

It is interesting to  observe that this alternative hypercharge
embedding appears also in the framework of heterotic  PS model
 \cite{ALR} where it can account for
the disappearance of fractionally charged particles. Of course in
the heterotic context non-standard hypercharge embeddings
are not useful for unification due to the
tight heterotic coupling relations.

\section{Proton stability, neutrino masses and all that \label{pd}}

One of the most serious problems of  SM extensions is proton decay.
In traditional GUTs it can be suppressed due to the  high unification scale.
However, such a suppression is not possible in low string scale models,
considered in the previous section.
In general there are three modes for proton decay (i) the gauge-mediated proton decay
(ii) the Higgs mediated and (iii)
higher dimension baryon number violating operators.
 In the PS model in particular, the exotic $SU(4)$ gauge bosons
$({{\bf3}},{\bf1},+\frac{2}{3})$, $({\bf\bar{3}},{\bf1},-\frac{2}{3})$ carry both baryonic
and leptonic quantum numbers  but they are known not to mediate
proton decay, due to the absence of di-quark coupling~\cite{PS,Lan}.
These particles can only contribute to  semi-leptonic processes, like
$\beta$-decay which leads to the bound $M_R > g_4 10^3 $GeV.

Higher dimension baryon number violating operators are expected to be
present in any GUT model embedded in string theory.
 They are suppressed by a factor $1/M_U^{d-4}$ where $d$ is the dimension of the
relative operator~\cite{Wein}.
In order to be safe with current proton decay limits, one has to prevent the
appearance of such operators up to a dimension as high as $d\sim 18$.
This suppression would look natural only in the case it could be associated with
a symmetry (gauged or global).

The standard PS model
contains $B-L$ as a gauged symmetry, but this is not enough to
avoid proton decay.  Already at sixth order,
$B-L$ conserving operators (e.g. $Q\,Q\,Q\,L$ originating from $F_L^4$)
lead to baryon number violation. Furthermore, the spontaneous breaking
of $B-L$ leads to additional operators suppressed only by $M_R/M_U$.
Fortunately,  the current ${U(4)}_C\times{U(2)}_L\times{U(2)}_R$
extension of the PS model, incorporates the required
$U(1)$ combination which corresponds to the baryon number itself.
Indeed, as can be seen from (\ref{fermions},\ref{ehiggs}),
$Q_C= 3\,B+L$ and thus
\ba
B &=& \frac{Q_C + Q_{B-L}}{4}
\label{Bsym}
\ea
is a global symmetry of the theory
(see discussion in the beginning of section \ref{sanom}),
which ensures the stability of the proton. Note that this symmetry
survives the PS breaking as the $\nu_H^c$ has zero baryon number.

There are general no-go theorems \cite{Banks:1988yz}
  against the survival
of  global symmetries in the context of string theory at least at
the perturbative level.
They are expected to be violated since black holes can absorb charged particles
but they cannot possess  global charges themselves, due to ``no-hair"
theorems.
However, there are arguments that in  the context of Type I and Type IIB
string vacua the assumptions of these no-go theorems can be evaded
as the Fayet--Iliopoulos term associated with the anomalous $U(1)$ can be set to
zero \cite{Ibanez:1999it}.
Moreover, global symmetries, as the baryon number,
are   expected to be violated due to non-perturbative phenomena,
 ( instantons).
Of course this  violation is expected to be suppressed and
 may not be sufficient for standard  baryogenesis scenarios.
However, in the brane-world models   
we can use some alternative higher-dimensional mechanisms  for the generation of
baryon asymmetry \cite{Dvali:1999gf}.

Higgs mediated operators are inversely proportional to the Higgs remnant
masses and could be dangerous for low string scale models. In the models
discussed here the  only Higgs light remnants are the triplets
$d^c_H$. These triplets are assigned with baryon  number under
(\ref{Bsym}), thus,
all their couplings  with ordinary matter are baryon conserving.
However, it is
desirable that $d^c_H$  triplet scalars  receive  masses
 (heavier than the proton)
since if they stay light enough, proton can still decay to
them (through baryon conserving processes). There are two possible
scenarios for generating masses for these scalars. The first is to
assume that the scalar potential -- the details of which are not
known since this is to be provided directly from string/brane
theory--  will eventually have some minimum which apart from
symmetry breaking could also provide ($M_R$) masses for these scalars.
The second is to introduce some extra scalar particles, namely the
triplets originating from the sextet
$D$ (see (\ref{sext})) which will mix
with  $d^c_H$ and thus provide masses for them.
We may assume a  scalar potential  of the form
\ba
{\cal V}&=&\rho^2\,D \,D^\dagger+\lambda\,\bar{H}\,\bar{H}\,D+
\mbox{c.c.}
\nonumber\\
&=&\rho^2\,({\tilde{d}^c}\,
{\tilde{d}^{c\dagger}}+{\tilde{d}}\,{\tilde{d}^{\dagger}})
+\lambda\,(d^c_H\,\tilde{d}\langle\nu^c_H\rangle+{d^c}_H^\dagger
{{\tilde{d}^{\dagger}}}\langle\nu^c_H\rangle^\dagger)+\cdots\label
{c1}
\ea
where $\rho$   and $\lambda$ are appropriate combinations of vevs.
In the case $c=1/2$ and in the lowest order,
$\rho^2=\langle H H^\dagger\rangle\sim M_R^2$, while
$\lambda =
\langle h_{Ri}h_{Rj}\rangle /M_U \sim M_{Z'}^2/M_R$.
Note that due to $SU(2)$ antisymmetry at least two different
$h_R$ fields are necessary in order to obtain a non-vanishing
coupling.
This superpotential provides triplet masses of the order
$m_1\sim M_Z^2/M_R$, $m_{2,3}\sim M_R$.
In the case $c=0$ one can assume similarly $\rho^2=\langle H H^\dagger\rangle$ and $\lambda =
\langle \eta\rangle$.

Baryon number is not the only global  symmetry  left from the anomalous
$U(1)$ breaking. As easily seen by the particle assignments
(\ref{fermions}-\ref{GUThiggs}) the lepton number corresponds to
the combination
\ba
{\cal L}=\frac{Q_C-3\,Q_{B-L}}{4}\label{ln}
\ea
In the case of  the baryon number all Higgs fields are neutral
under it and the symmetry remains exact at the   perturbative theory level.
On the contrary the $ \nu^c_H$ has lepton number
(although $h, h_R, \eta$ are neutral) and it
will thus break ${\cal L}$ spontaneously
and give rise to a massless Goldstone boson.
One possible solution to this problem is discussed in \cite{BU1} where a
deviation from the orientifold point (along a direction that conserves
baryon number) is considered.
Furthermore, one may note that the correct lepton number for all
fermions and electroweak Higgs fields is reproduced by a more
general formula
$
{\cal L}'=k\,Q_C-\frac{3}{4}\,Q_{B-L}+\left(k-\frac{1}{4}\right)\,(Q_R-\,Q_L)
\label{gln}
$
where $k$ is an arbitrary number,
and (\ref{ln}) corresponds to the particular case $k=\frac{1}{4}$.
This alternative definition preserves the fermion charges but
can give different PS Higgs charges. In the case
$c=1/2$ and  choosing $k=0$ we have
\ba
{\cal L}'=-\frac{3}{4}\,Q_{B-L}-\frac{1}{4}\,Q_R+\frac{1}{4}\,Q_L
\ea
which renders $\nu^c_H$ neutral.
Thus, lepton number is not broken at the level
of PS symmetry ($M_R$), but at the $M_{Z'}$ scale as the
right-handed doublets ($h_R$), utilized in this case for the additional
$U(1)$ breaking, are charged. This leads to the interesting
possibility that the lepton number breaking is associated to the
breaking of an additional abelian symmetry.

Apart from the low energy values of the Weinberg angle and the strong
coupling,
a consistent string model is also expected to reproduce the
low energy  fermion mass pattern.
The  PS symmetry implies unification
of all Yukawa couplings. Thus for the heaviest generation,
which is expected to receive mass at tree-level,
we have  $m_{\tau}=m_{b}$ at the brane scale.
In an ordinary GUT, the observed low energy
difference of the two running masses is attributed to the
$SU(3)$-contributions in   $m_b$.
In low energy
unified models the range $M_U-M_Z$ is too short to account for
the $m_b-m_{\tau}$ difference, however,  the required enhancement can be
anticipated by the ratio of the gauge couplings given in (\ref{ratio}).
In addition, the rest of the fermion masses and mixings  are expected to
be easily reproduced due to the potential presence of extra Higgs doublets
(which as shown above do not affect the string scale)
and  generation mixing.

{}For neutrino masses in particular, recent experimental  explorations
have shown that it is likely that a  crucial role is played by
 the right-handed neutrino $\nu^c$ which is absent in the SM.
In most extensions of the SM theory, $\nu^c$ receives a large
mass of the order of the unification scale. Then,  the
see-saw mechanism is used to generate a tiny mass for the left-handed neutrino,
which is compatible with experimental and astrophysical
limits.
In the context of a  D-brane approach to SM
one has to assume that $\nu^c$ will possibly arise as a gauged
neutral fermion propagating in the bulk and explain the
light neutrino mass by the smallness of the brane-bulk
couplings, naturally suppressed by the
bulk volume \cite{bulknu,bulknulr,bulknur}.
 On the contrary, one important feature of the
PS extension of the SM model (and left-right models in general),
 is that the right-handed neutrino lives on the brane as any other
fermion of the SM.
In addition, a Dirac neutrino mass term $L\,\nu^c\,\langle H_u\rangle$
is generated by
the coupling  $F_L\,\bar{F}_R\,h$ which cannot be forbidden as it also
generates masses for all the SM fermions.
A Majorana mass is also possible from an effective term $\kappa\,F_R\,F_R$
where $\kappa$ an appropriate vev combination. These terms lead
to the  neutrino mass matrix
\ba
m_\nu=
\bordermatrix{&{\nu}&\nu^c\cr
\nu&0&\langle H_u\rangle\cr
\nu^c&\langle H_u\rangle&\kappa\cr}
\ea
with eigenvalues $m_{light}\sim \frac{\langle H_u\rangle^2}{\kappa},
m_{heavy}\sim \kappa$ assuming
$\kappa >\langle H_u\rangle$.
 For
the $c=0$ model the simplest choice is
$\kappa= \frac{\langle H^\dagger\,H^\dagger\rangle}{M_U}=\frac{M_R^2}{M_U}$ which gives  adequately suppressed neutrino masses for
$M_R\lsim M_U\sim 10^{10}$ (see table \ref{czunf}).
For the $c=1/2$ model  $\kappa=\langle H^\dagger\,H^\dagger\, h_R\,h_R\rangle/M_U^3$ requires
$M_U>10^8$ in order to suppress enough
the left-handed Majorana neutrino  masses at an experimentally
acceptable range.
Hence, in this case  a different mass generation mechanism
must be employed. A possible solution applicable in general left-right symmetric
model has been presented in \cite{bulknulr}.
The main idea is to consider a bulk right-handed neutrino that mixes
only with the brane right-handed  neutrino. An additional
possibility would be to consider
 masses for the bulk neutrinos along the lines proposed in \cite{bulknur},
as well as potentially unsuppressed gravitational matter interactions \cite{MF},
and utilize a generalized see-saw mechanism (including the Kaluza--Klein
excitations of bulk neutrinos) to reconcile the experimentally acceptable
neutrino masses with a low string scale.

\section{Conclusions}

In this paper we have  explored a generic
Pati--Salam like model based on an ${U(4)}_C\times U(2)_L\times U(2)_R$
gauge symmetry,  compatible with  a D-brane configuration.
We have  found two consistent models
one with the standard and one with an alternative
hypercharge embedding.
The former is  compatible with the low energy data for an intermediate
string scale of the order of  $10^{10}$ GeV, while
the later  is shown to   be compatible with the electroweak data
for a string scale  of the order of {\it 5-7} TeV provided that the
${U(4)}_C$ and ${U(2)}_R$ brane sets have equal couplings ($\alpha_4=\alpha_R$) while the
${U(2)}_L$ coupling  is about a half of this value ($\alpha_L\sim\alpha_4/2$).

Both scenarios contain an extra abelian factor which can break
at an acceptable  scale by vevs of appropriate
scalar fields incorporated in the models.
In the low string scale case we have identified lepton number with
a global symmetry of the theory whose breaking is associated
with the breaking of the additional abelian factor.

Proton stability is assured, as an anomalous combination of the surplus
abelian factors of the original gauge group is identified with
the baryon number.
This combination is to be broken by a generalized Green--Schwarz mechanism at
the  string level leaving behind baryon number as an exact
global  symmetry.

 The right-handed neutrino is part of the non-trivial fermionic representations
of the theory, while there can exist mechanisms which
make the left-handed Majorana mass compatible with
recent data.  More particularly, in the case of intermediate string scale
the lightness of the
 neutrino can be guaranteed by a see-saw mechanism at the
brane level while in the case of a low energy string scale a generalized
see-saw mechanism incorporating bulk sterile neutrinos and possibly
bulk masses is required.

It would be interesting if the model presented here, and especially
the variation with low string/brane scale, could find a direct
realization in the  context of  Type I  constructions \cite{PSS}.

\section*{Acknowledgements}
The work is partially supported by the European Commission
under the TMR contract ERBFMRX-CT96-0090 and RTN contract
 HPRN-CT-2000-00152.

\newpage


\end{document}